*Communication*

# Ultra-Broadband and Compact 2 × 2 3-dB Silicon Adiabatic Coupler Based on Supermode-Injected Adjoint Shape Optimization

Hongliang Chen [1,2], Guangchen Su [1,2], Xin Fu [1] and Lin Yang [1,2,*]

1 State Key Laboratory on Integrated Optoelectronics, Institute of Semiconductors, Chinese Academy of Sciences, Beijing 100083, China; chenhongliang@semi.ac.cn (H.C.); suguangchen@semi.ac.cn (G.S.); fuxin@semi.ac.cn (X.F.)
2 College of Materials Science and Opto-Electronic Technology, University of Chinese Academy of Sciences, Beijing 100083, China
* Correspondence: oip@semi.ac.cn

**Abstract:** The 2 × 2 3-dB couplers are one of the most widely used and important components in silicon photonics. Here, we propose an ultra-broadband and compact 2 × 2 3-dB adiabatic coupler defined by b-splines and optimized with an efficient supermode-injected adjoint shape optimization. By employing mode adiabatic evolution and mode coupling at two different wavelength ranges, respectively, we achieve an ultra-broad bandwidth of 530 nm from 1150 nm to 1680 nm with a power imbalance below ±0.76 dB in a compact coupling length of 30 μm according to our simulation results. The supermode-injected adjoint shape optimization can also be applied to the design of other photonic devices based on supermode manipulation.

## 1. Introduction

With the development of artificial intelligence (AI) technology, high-speed, and low-latency datacom interconnects are required for the training and inference of AI large models on high-performance computing (HPC) clusters. Silicon photonics is a promising solution for its high integration density, low cost, and CMOS process compatibility [1]. The 2 × 2 3-dB couplers are one of the most widely used and important components in silicon photonics. Conventional 3-dB couplers based on directional couplers (DCs) have limited bandwidth [2]. Couplers based on Multimode interference (MMI) offer a broad bandwidth but at the cost of considerable loss and a large footprint [3]. There have been several approaches to enhance the performance of the 3-dB couplers, including bent DCs [4,5], subwavelength grating (SWG) assisted DCs [6,7], and adiabatic couplers [8–10]. Bent DCs can operate at wide bandwidths by achieving phase matching between two waveguides with different propagation constants [4,5]. This requires precise control of the waveguide and gap widths. SWG structure has been used to tailor the dispersion of waveguides, achieving broad bandwidths in compact lengths. An ultrabroad bandwidth of 270nm from 1400nm to 1670nm in a total length of 24.4 μm in SWG-assisted DCs is demonstrated, but it is designed in a minimum feature size of 64 nm, which may be challenging to fabricate[6]. Adiabatic couplers often require a long length to achieve a broad bandwidth. A variety of adiabatic curves have been proposed to realize fast adiabaticity and reduce the coupling length [9–12]. A short coupling length of 11.7 μm can be achieved in an adiabatic coupler optimized by the FAQUAD protocol. However, this comes at the price of a limited bandwidth of 75 nm [10]. Therefore, achieving broadband 2 × 2 3-dB adiabatic couplers with compact footprints is still demanding.

In recent years, there has been a growing interest in inverse design, which has resulted in the design of various devices, such as mode manipulation devices [13,14], wavelength (de)multiplexers [15,16], grating couplers [17], and micro resonators [18]. In addition, 2 × 2 3-dB bent couplers with low loss and compact footprint can be realized by inverse design [19,20]. Adjoint shape optimization is a gradient-based inverse design and is suitable for the design of an initial structure. Meanwhile, it is also easy to apply shape constraints, leading to fabrication-friendly devices such as 1 × 2 splitters [21], waveguide crossing [22], and polarization rotators [23]. However, there have been limited investigations into the utilization of adjoint shape optimization for the design of adiabatic couplers.

In this work, we propose an ultra-broadband and compact 2 × 2 3-dB adiabatic coupler working for TE polarization based on silicon strip waveguides of the SOI platform. The coupler, with a minimum feature size of 100 nm, can be fabricated using electron beam lithography. By optimizing the shape and length of the coupling region, the adiabatic coupler can operate in mode evolution at longer wavelengths and mode coupling at shorter wavelengths, leading to an expansion of the bandwidths. The coupling region defined by b-splines is optimized efficiently by a novel supermode-injected adjoint shape optimization, which uses supermode as the forward and adjoint sources, thus reducing the optimization area and the simulation time. The simulation result indicates that we can achieve an ultrabroad bandwidth of 530 nm from 1150 nm to 1680 nm with a power imbalance under 0.76 dB at a compact coupling length of 30 μm.

## 2. Design and Principle

Figure 1a shows a three-dimensional view of the device, which is based on the SOI platform with a 220 nm top layer of silicon, a 3 μm buried oxide layer, and a 2 μm silicon dioxide cladding. The device consists of three regions as shown in Figure 1a.

In Region I, WG1 and WG2 of widths $W_1 = 380$ nm and $W_2 = 500$ nm, respectively, are brought closer via two S-bends of length $L_1$, resulting in a decrease of the gap between them from $G_0$ to $G_1$ and forming a dual waveguide system. After Region I, only one supermode of the dual waveguide system is excited and enters into Region II. In Region II, the supermode achieves adiabatic evolution while the widths of WG1 and WG2 slowly vary to reach the same $W_3 = 410$ nm, resulting in an equal distribution of optical power into the two waveguides. In Region III, the gap between them is increased from $G_1$ to $G_2$ by using two S-bends of length $L_3$. $G_0$ and $G_2$ should be large to prevent excitation of modes in the other waveguide. Therefore, we choose $G_0 = 1.6$ μm and $G_2 = 1.65$ μm. $L_1$ and $L_2$ should be sufficiently long to deteriorate the coupling of the supermodes. Thus, we choose $L_1 = 30$ μm and $L_2 = 12$ μm.

W1 and W2 should have a significant difference to prevent modes coupling in the input region (Region I). If this difference is too small, then the required length of the input S-bend will be correspondingly longer. W3 corresponds to the same width that the WG1 & 2 reach at the end of the coupling region, and it impacts the magnitude of the width variation of the two waveguides at the coupling region (Region 2). The choice of W3 is not unique, but it should be considered that the narrow waveguide (WG1) is more sensitive to the change in width. Therefore, the narrow waveguide should have a lesser width variation, while the wider waveguide (WG2) should have a larger width variation. Consequently, we choose the width variation W2−W3 of the wide waveguide to be three times the width variation W3−W1 of the narrow waveguide.

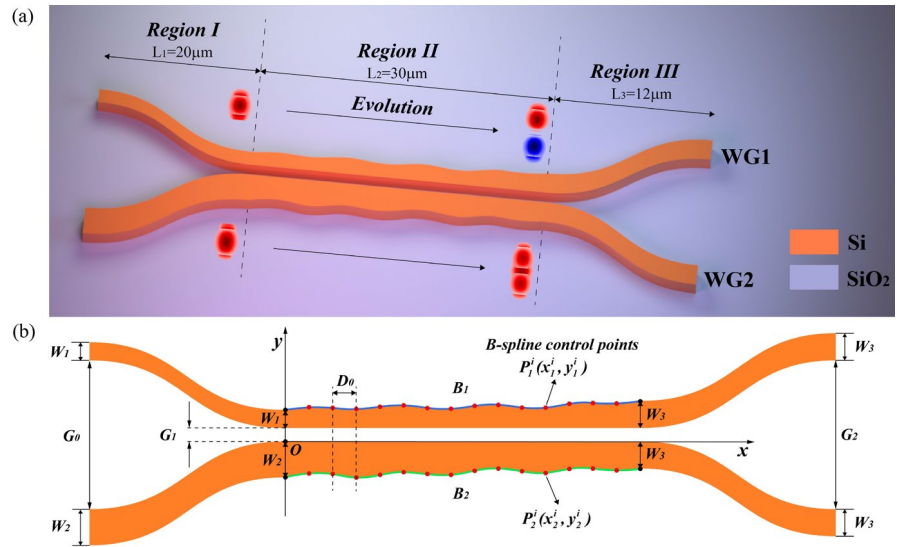

**Figure 1.** (**a**) 3D Schematic of the adiabatic coupler and the supermodes at the beginning and the end of Region II. (**b**) Top view of the adiabatic coupler.

In Region II, WG1's lower edge and WG2's upper edge are straight and aligned with the x-axis, which is also the propagation direction, and the gap between them is $G_1 = 100$ nm to satisfy the actual fabrication process requirements. The y-axis is located at the left end of Region II. Meanwhile, the optimization of the upper edge shape in WG1 ($\boldsymbol{B_1}$) and lower edge shape in WG2 ($\boldsymbol{B_2}$) is to be carried out using two fourth order clamped b-splines, defined by their control points as follows:

$$\boldsymbol{B_1}(t) = \sum_{i=0}^{n} N_{i,p}(t) \boldsymbol{P_1^i} \qquad (1)$$

$$\boldsymbol{B_2}(t) = \sum_{i=0}^{n} N_{i,p}(t) \boldsymbol{P_2^i} \qquad (2)$$

where $\boldsymbol{P_1^i}, \boldsymbol{P_2^i}$ represents the $(i+1)$th control point of $\boldsymbol{B_1}(t), \boldsymbol{B_2}(t)$. The coordinates of $\boldsymbol{P_1^i}$ are $x_1^i$ and $y_1^i$ and the coordinates of $\boldsymbol{P_2^i}$ are $x_2^i$ and $y_2^i$. The $N_{i,3}(t)$ denotes the B-spline basic functions of degree three. This indicates that each point on the curve is a linearly weighted sum of the four closest control points. Low order b-splines may produce sharp shapes, which can pose difficulties in fabrication and deteriorate fabrication tolerances. On the other hand, higher order b-splines exhibit less flexibility, causing them to be more rigid and difficult to optimize. Fourth order b-splines are suitable, as they maintain both flexibility and smoothness. The curve's control points are distributed evenly along the x-axis with a large distance of $D_0 = 2$ μm between adjacent control points, ensuring a smooth and flat curve, resulting in a length of $L_2 = 30$ μm containing 16 control points. To reduce the number of optimization parameters, the x coordinates of each control point are fixed, while the y coordinates of $\boldsymbol{B_1}$ are proportional to those of $\boldsymbol{B_2}$, as specified by:

$$y_1^i = \frac{W_3 - W_1}{W_2 - W_3}(y_2^i + W_2) + (W_1 + G_1) \qquad (3)$$

To guarantee a seamless connection between the regions, both endpoints of $\boldsymbol{B_1}$ and $\boldsymbol{B_2}$ will be firmly fixed to eliminate any discontinuities. Thus, the total number of optimization parameters is 14.

Region II aims to guarantee supermode adiabatic evolution while avoiding any coupling with other supermodes during geometric changes. This is similar to the principle of multimode waveguide bend, which prevents inter-mode crosstalk when the geometry undergoes bending. Therefore, adapting shape optimization to an adiabatic coupler only requires replacing the sources from the single waveguide system's modes with the dual waveguide system's supermodes. We defined the FOM (figure of merit) as follows:

$$FOM = T_{odd\ supermode} \quad (4)$$

where $T_{odd\ supermode}$ is the average transmission of the odd supermode only through Region II in the wavelength range from 1290 nm to 1630 nm with sampling every 10 nm. In Region II, the main physical process involves the evolution of the supermodes, resulting in a slower change in the shape of the coupler. As a result, the system exhibits almost lossless characteristics, thus the main factor that deteriorates the FOM is the inter-supermode coupling rather than the total loss. Since there are only two supermodes in the dual waveguide system, if one supermode's transmission is high, then the transmission of the other supermode should also be high when light is launched from another input port. Actually, according to our simulation during optimization, the transmission of both the odd and even supermodes is always nearly identical. However, the odd supermode performs slightly worse than the even supermode since WG1, which has a narrower width, is more sensitive to geometry variations. Therefore, we simply select the transmission of the odd supermode as the FOM and thus cut the simulation time in half.

We use 3D-FDTD to obtain the FOM. In every iteration, we employ the adjoint method to calculate the gradient. The adjoint shape optimization was first applied to electromagnetic design by Lalau-Keraly et al. [21]. According to the adjoint method, the derivative of the FOM with respect to dielectric permittivity at every point in the design region can be expressed as follows:

$$\frac{\partial FOM}{\partial \varepsilon(x)} = Re[\boldsymbol{E}_{fwd}(\boldsymbol{x}) \cdot \boldsymbol{E}_{adj}(\boldsymbol{x})] \quad (5)$$

Here, $\boldsymbol{E}_{fwd}(\boldsymbol{x})$ is the electric field of forward simulation that can be used to obtain the FOM. The $\boldsymbol{E}_{adj}(\boldsymbol{x})$ is the electric field of adjoint simulation, which consists of sending the desired mode backward into the coupler. For a complete derivation and a more detailed study of the adjoint method, we refer to [24]. As stated earlier, we defined the FOM as the transmission of the odd supermode, therefore we send the supermode backward to the coupler in the adjoint simulation. By changing the adjoint source to supermode, this supermode-injected technique enables the application of adjoint shape optimization to the design of supermode manipulation devices. We then differenced each optimization parameter and calculated the resulting change in the distribution of permittivity. By combining this and the derivative of the FOM with respect to permittivity, we can obtain the FOM gradient with respect to the optimization parameters

The L-BFGS-B algorithm is subsequently utilized to update the parameters based on the acquired FOM and gradients. The forward source is set to be the odd supermode at the beginning of Region II, as shown in Figure 2a, while the adjoint source is set to be the same odd supermode at the end of Region II, as shown in Figure 2b. The even supermodes located at the same position as the odd supermodes are also illustrated in Figure 2c–d.

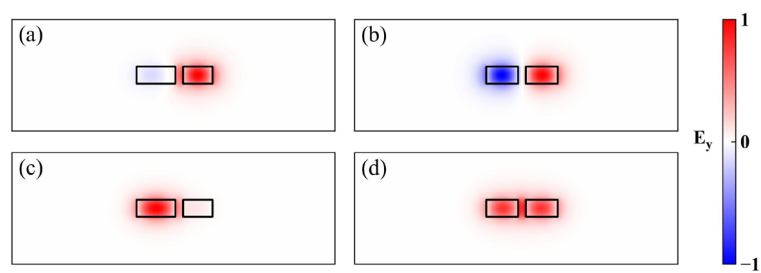

**Figure 2.** The profile of (**a**,**b**) odd and (**c**,**d**) even supermode at the input and output of region II.

## 3. Simulation Results

After 28 iterations, the figure of merit (FOM) reached 0.9966, as depicted in Figure 3b. The iteration process takes 10 h on a server equipped with dual 2.90-GHz Intel Xeon Platinum 8268 CPUs and 256 GB RAM. The optimized $B_2$ shape is depicted as the red solid line in Figure 3a, while the conventional linear taper shape is represented in a blue dash line for comparison in the same figure.

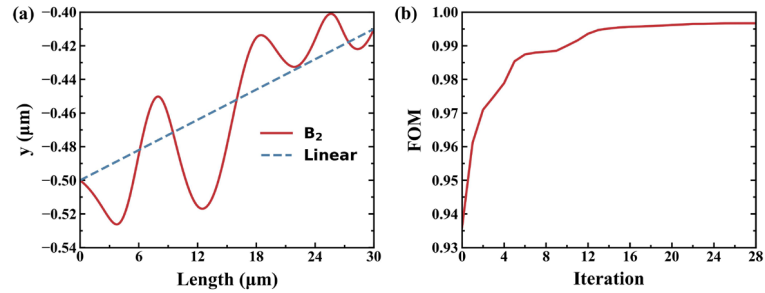

**Figure 3.** (**a**) The shape of B2 in solid line and conventional linear taper shape in dash line for comparison. (**b**) Plot of the figures of merit with the increasing iterations.

The device's transmission from the input port in Region I to the bar and cross output port in Region III is illustrated in Figure 4a covering the wavelength range of 1150 nm–1680 nm. For the linear design, the transmission of bar and cross port is approximately −2 dB and −4 dB over the wavelength range of 1400 nm–1650 nm, with an obvious degradation at shorter wavelengths below 1350 nm. Conversely, for the optimized design, the power imbalance stands at 3 ± 0.5 dB in a broad bandwidth of 204 nm from 1474 nm to 1678 nm, and 3 ± 0.76 dB over the entire range of 1150 nm–1680 nm.

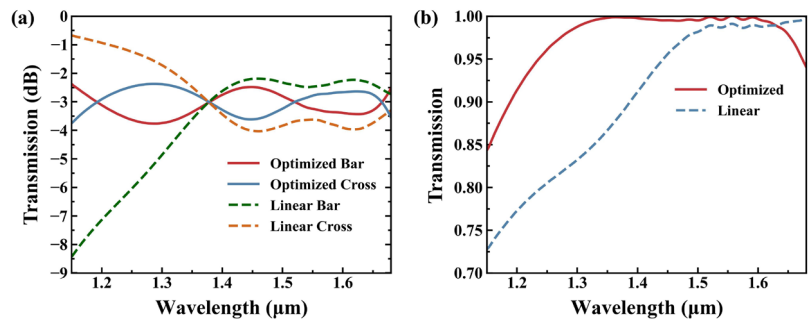

**Figure 4.** (**a**) Simulated power splitting ratio of optimized design and linear design. (**b**) Transmission for odd supermode through optimized design and linear design.

Figure 4b illustrates the odd supermode's transmission through Region II. The data clearly indicates that the optimized design performs better than the linear design. However, the transmission of shorter wavelengths remains low when compared to the transmission of longer wavelengths. Nonetheless, the power splitting ratio of shorter wavelengths does not decrease significantly compared to the linear design. This is because the splitting ratio depends not only on the magnitude of the supermodes' transmission but also on the phase difference between odd and even supermodes at the end of Region II, which can be expressed as follows:

$$\frac{T_{bar}}{T_{cross}} = \frac{T_{odd} + 2\sqrt{T_{odd}T_{even}}\cos\Delta\varphi + T_{even}}{T_{odd} - 2\sqrt{T_{odd}T_{even}}\cos\Delta\varphi + T_{even}} \quad (6)$$

Here, $T_{odd}$ is the transmission of the odd supermode, while $T_{even}$ refers to the transmission of the even supermode. $\Delta\varphi$ represents the phase difference between the odd and even supermodes at the end of Region II. Additionally, $T_{bar}$ denotes the transmission at the bar port, and $T_{cross}$ the transmission at the cross port. When either $T_{even}$ or $T_{odd}$ is sufficiently small, the power splitting ratio $T_{bar}/T_{cross}$ approaches 1, indicating the coupler's adiabatic operational status. When the phase difference $\Delta\varphi$ equals $(2n + 1)\pi/2$, the power splitting ratio $T_{bar}/T_{cross}$ is also equal to 1, indicating that the device is operating as a 3-dB directional coupler. Thus, at longer wavelength range, the device is optimized for realizing supermode adiabatic evolution. At shorter wavelengths below the FOM's range, the device operates as a directional coupler by using the appropriate coupling length $L_2 = 30$ μm to achieve the $(2n + 1)\pi/2$ phase difference between odd and even supermodes. Figure 5a illustrates the phase difference between odd and even supermodes at the end of Region II. It can be noted that there is a flat band located around $\pi/2$ in the short wavelengths, which is consistent with our analysis and Equation (6) above.

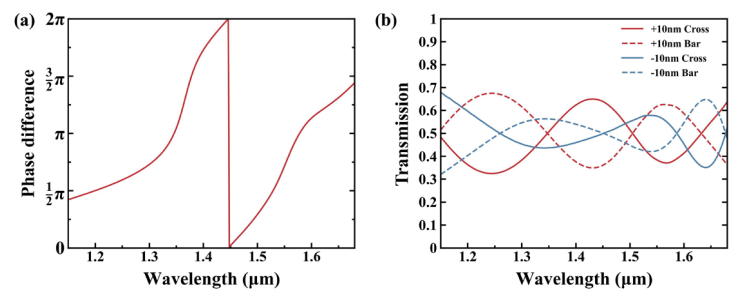

**Figure 5.** (**a**) The phase difference between odd and even supermodes at the end of Region II. (**b**) Simulated transmission for a waveguide width deviation of $\Delta w = \pm 10$ nm.

In order to analyze the fabrication tolerances, we simulate the transmission for a waveguide width deviation of $\Delta w = \pm 10$ nm, while keeping a constant center-to-center distance for the two waveguides. As Figure 5b shows, the splitting ratio degraded to 50% ± 18%, indicating that the coupler is sensitive to geometry deviations. The coupler has the same sensitivity as conventional DC, as there is the same degradation both in the short wavelength range operating as a conventional DC and the long wavelength range operating as an adiabatic coupler. To achieve a more robust design, we can increase the gap and the waveguide width, which may lead to a longer device length, or add a robust term to the FOM [23].

Figure 6a–f show the simulated light propagation profiles at the wavelengths of 1.2 μm, 1.4 μm and 1.6 μm when the light is launched from the WG1 and WG2 input ports, respectively. At all three wavelengths, it can be seen that the light is evenly split into the two output ports.

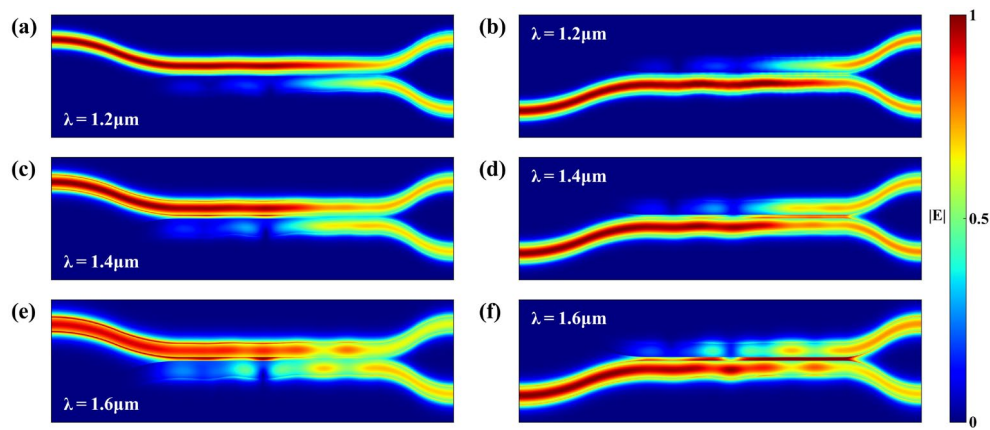

**Figure 6.** Simulated light propagation profiles for the 2 × 2 3-dB adiabatic coupler at wavelengths of (**a**,**b**) 1.2 μm, (**c**,**d**) 1.4 μm, (**e**,**f**) 1.6 μm with TE0 mode input from the upper and lower ports, respectively.

## 4. Conclusions

In conclusion, we have proposed an ultra-broadband and compact 2 × 2 3-dB adiabatic coupler on SOI. We adapted the adjoint shape optimization for the design of an adiabatic coupler by using supermodes as the forward and adjoint source. An ultra-broad bandwidth is achieved in a short length by combining the mode adiabatic evolution and mode coupling at different wavelength ranges, respectively. The simulation result shows the power imbalance of the 3-dB coupler is under ±0.76 dB over an ultra-broad bandwidth of 530 nm from 1150 nm to 1680 nm with coupling length being 30 μm and total length being 62 μm. We believe our device can find numerous applications in integrated photonics. The supermode-injected adjoint shape optimization can also be applied to other photonic devices based on supermode manipulation, resulting in a more compact device footprint.

**Author Contributions:** Conceptualization, H.C.; methodology, H.C. and G.S.; software, H.C.; validation, H.C.; formal analysis, H.C.; investigation, H.C.; resources, H.C.; data curation, H.C.; writing—original draft preparation, H.C.; writing—review and editing, H.C. and X.F.; visualization, H.C.; supervision, L.Y.; project administration, X.F. and L.Y.; funding acquisition, X.F. and L.Y.; All authors have read and agreed to the published version of the manuscript.

**Funding:** This research was funded by The Strategic Priority Research Program of Chinese Academy of Sciences under Grant XDB43000000, in part by the National Science Fund for Distinguished Young Scholars under Grant 61825504, in part by the National Natural Science Foundation of China under Grant 61975198.

**Institutional Review Board Statement:** Not applicable.

**Informed Consent Statement:** Not applicable.

**Data Availability Statement:** Relevant data are available from the authors upon reasonable request.

**Conflicts of Interest:** The authors declare no conflicts of interest.